\documentclass[sn-mathphys-num]{sn-jnl}

\usepackage{graphicx}%
\usepackage{amsmath,amssymb,amsfonts}%
\usepackage{xcolor}%

\newcommand{\ket}[1]{ \left| #1 \right> }
\newcommand{\bra}[1]{ \left< #1 \right| }

\newcommand{\Sch}{Schr\"{o}dinger}

\newcommand{\be}{\begin{eqnarray}}
\newcommand{\ee}{\end{eqnarray}}

\newcommand{\up}{\uparrow}
\newcommand{\dn}{\downarrow}
\newcommand{\Up}{\Uparrow}
\newcommand{\Dn}{\Downarrow}
\newcommand{\vh}{{\bf h}}

\newcommand{\vt}{{\bf t}}

\newcommand{\sthird}{\mbox{\small $\sqrt{ \frac{1}{3}}$ }}
\newcommand{\stthird}{\mbox{\small $\sqrt{ \frac{2}{3}}$ }}
\newcommand{\stwo}{\mbox{\small $\sqrt{ \frac{1}{2}}$ }}
\newcommand{\sqf}[1]{\mbox{\small $\sqrt{ \frac{1}{#1}}$ }}

\newcommand{\myfig}[2]
{ \centerline{\resizebox{!}{#1\textwidth}{\includegraphics{#2}}} }

\begin{document}

\title[Extended Wigner's friend]{The extended Wigner's friend, many- and single-worlds and reasoning from observation}

\author*[1,2]{\fnm{Andrew} \sur{Steane}}\email{a.steane@physics.ox.ac.uk}

\affil*[1]{\orgdiv{Department of Atomic and Laser Physics}, \orgname{University of Oxford}, \orgaddress{\street{Clarendon Laboratory, Parks Road}, \city{Oxford}, \postcode{OX1 3PU}, \state{}, \country{England}}}

\abstract{
The concept of an isolated system, and
Frauchiger and Renner's extended `Wigner's friend' scenario are discussed.
It is argued that: 
(i) it is questionable whether the approximation of the isolated system
is valid when measurement-like processes are involved;
(ii) one may infer, from Frauchiger and Renner's thought-experiment,
and similar thought-experiments, that any interpretation of 
quantum theory involving {\em subjective collapse} fails;
(iii) this does not distinguish single-world from many-world
(relative-state) interpretations of quantum theory;
(iv) reasoning from observations has to take
into account the possible quantum-erasure of those observations 
if it is to be valid reasoning;
(v) a single-world interpretation is valid if certain kinds of outcome
are not quantum-erased in the future. 
}

\keywords{Isolated system, Wigner's friend, quantum measurement, interpretation of quantum mechanics}

\maketitle

\section{Introduction}

The relationship between physical observations and the mathematical apparatus 
of quantum mechanics has been discussed for over one hundred years, and has
given rise to a large literature. An important thought-experiment is that
of {\em \Sch's cat} \cite{PeresBk1993}. This thought-experiment 
brings into focus questions about knowledge, observation and measurement,
and the relationship between the physical world and the mathematical
apparatus of quantum theory. In these respects it is a useful 
thought-experiment. However it has not proved possible to settle disputes about
quantum foundations by appeal to this thought-experiment alone. 

A related thought-experiment is that of {\em Wigner's friend} \cite{Wigner1967,PeresBk1993}. Recently some
extensions of the Wigner's friend scenario have been described, with a view to
making it possible to make assertions about quantum theory in a more concrete
way. For example, Frauchiger and Renner (FR) put forward three assertions concerning
quantum theory and physical phenomena and argued, from analysis of their
thought-experiment, that the three assertions cannot all be true \cite{Frauchiger2018}. This is helpful because it gives us something
specific to reason about. I will argue in this paper that 
Frauchiger and Renner's proposed list did not sufficiently consider two issues: first that the approximation of an isolated system may not be warranted when
large systems with chaotic dynamics are involved, and secondly that some
physical processes leave permanent records. 
I will argue that the physical notion of a {\em permanent record} (to be 
defined), combined with the standard mathematical apparatus of quantum field 
theory, together constitute sufficient conditions for 
the validity of a single-world interpretation of quantum mechanics.
I will also argue that reasoning from observation has to take into account
the possibility of such a record.

The paper proceeds as follows. Section~\ref{s.isolated} discusses the notion of 
an {\em isolated} system, pointing out that it is not applicable in particular 
cases. 
Section~\ref{s.singlemw} summarises the difference between single-world and 
many-world (or relative-state) quantum theories.
Section~\ref{s.project} gives a standard mathematical treatment of
measurement, drawing attention to the physical implications of a
permanent record. Section~\ref{s.extendW} summarizes existing
responses to the extended Wigner's friend (EWF) thought experiment. Section \ref{s.FR} describes
that thought-experiment in detail and presents the argument of FR. We then
show how the reasoning in this scenario depends on what kind of process
takes place at the step called `observation'. If it is a process that can be reversed
by quantum erasure then one line of argument applies; if it is not then another
line of argument applies. Section \ref{s.singlerev} revisits the distinction
between single-world and many-world quantum theories in light of what has been discussed.
Section~\ref{s.conclude} concludes.

\section{The method of the isolated system}  \label{s.isolated}

When, in developing physical reasoning, we invoke the notion of an
{\em isolated system}, this is often a legitimate thing to do, but one should not forget that it is an idealization. No part of the physical world is entirely isolated from other parts in fact. The idealization is valuable when it provides a good approximation to the evolution over time of those aspects 
of some system in which one is taking an interest. However it is not self-evident that the approximation of the isolated system will ever hold for experiments of the \Sch\ cat type. 

Before discussing this, let us briefly comment on the question, whether the physical universe as a whole is isolated. Quantum physics is unable to answer that question because any `answer' it gives must in fact be a form of circular reasoning. What we can reasonably expect from quantum physics is guidance on how parts of the physical world influence one another.

In the following I will show that in order to make the claim that
it is possible to attain stable quantum interference of states of a cat,
or of a wide range of states of large chaotic systems, one must make strong
assumptions about areas of physics, such as quantum gravity, for which no
agreed formalism exists, and one must trust that established physics attains
a degree of precision in its description of physical phenomena far beyond 
that which has been tested empirically, and one must make implausible claims about gravitational shielding and noise compensation. This point is not new but it is apt to be overlooked; the discussion in this section reviews it and draws attention to some relevant recent work.

Consider now the kind of physical process that is involved in discussions of
measurement and observation. These involve an interaction between two quantum systems, one of which is large and realises chaotic motion in its internal dynamics. By {\em large} here we mean that amounts of mass, length and time similar to those involved in the behaviour of an ordinary household cat
are involved. By {\em chaotic} we mean that some aspects of the system
exhibit extreme sensitivity to perturbations and extremely complicated motions. For example, this is true of the motion of gas molecules owing to their collisions, and of the libration (e.g. tumbling) of water molecules in liquid state such as in the body of a cat.
It is also true of many motions in ordinary solids and liquids at room
temperature, and of the sequences of chemical reactions taking place in living things. In such cases the conditions that are required in order to justify the approximation of the isolated system are extreme, as we now discuss.

Consider, for example, a Schr\"{o}dinger cat type of experiment involving quantum superposition of a mass $m=4\,$kg (the mass of an ordinary cat) in two locations separated by $\Delta x=10\,$cm for a time $t=1\,$s. 
Let $g$ be the gravitational acceleration in the direction of the separation
(or, equivalently, the acceleration, relative to a local inertial frame, of the apparatus giving rise to, or responding to, the superposition). After time $t$ a phase 
$m g \Delta x\, t/\hbar$ will accumulate between
the two parts of the superposition. 
This phase changes by $\pi$ when $g$ changes by $\Delta g \simeq 10^{-33}\,{\rm ms}^{-2}$. This indicates the degree to which $g$ will have to be known or controlled at timescales of order 1 second if evidence of superposition is to be obtained. 
The motion of gas particles at standard temperature and pressure is, meanwhile, very much more sensitive still, owing to the amplifying effects of collisions.
By using Eqn (1) of \cite{Steane2024a} one finds that
if $g$ changes by $10^{-100}\,{\rm ms}^{-2}$ then motion in a classical hard sphere gas at standard temperature and pressure
is substantially affected after about 50 collision times, which happens in about ten nanoseconds. A one nanometre displacement of a hydrogen atom on the other side of the Milky Way galaxy, for example, would be sufficient to disturb the gas. In the quantum treatment, the randomizing effect of uncontrolled such changes in $g$ takes the form of decoherence \cite{Goussev2012,Yan2020,Peres1984}.
\label{key}
The effect of gravitational noise at higher frequencies is given in \cite{Steane2017}. The stochastic gravitational background will make the interference unobservable unless it is somehow suppressed or compensated-for.

The cat's mass of $4\,$kg will itself produce gravitational acceleration of
$3\times 10^{-8}\,{\rm ms}^{-2}$ at a distance of $10\,$cm. The cat may develop an entanglement with the local gravitational field, though the theory of this is as yet undeveloped. Any instrument able to stabilize the field will need either to shield field changes in a passive manner, which would presumably require a huge mass and density, or else it would need to
be as sensitive to the field as the cat is. Such an instrument would also couple to the gravitational field of the cat itself, so would need to be included in the `system' (the contents of the notional box) to which the approximation of isolation is being applied. It is simply unknown whether the physical laws which describe such systems can in principle allow the required outcome (a superposition with a stable interference phase) to be achieved.

Similar remarks can be made about magnetic and electric field noise and the impact of thermal radiation. For a chaotic system the influence of Unruh radiation may be especially hard to avoid or control.
When molecules collide, they accelerate and emit Unruh radiation and consequently undergo momentum diffusion; the effect is sufficient to randomise their motion in a few nanoseconds at ordinary temperature and
pressure \cite{Steane2024a}.
All this is before one has contemplated the task of somehow reversing or otherwise manipulating in detail the motion of $10^{26}$ water molecules, and similar numbers of gas molecules and infrared photons, all in the
presence of gravitational noise. 

It is well known that complicated and sensitive behaviours 
are involved in experiments of this kind. However, sometimes these issues
are referred to using the phrase `technical difficulties' or as requiring merely 
`technological progress' \cite{Frauchiger2018}.
That is an incorrect use of language
because it assumes knowledge that we do not have. 
There are two difficulties. First, the Standard Model of quantum field theory is almost certainly incomplete or incorrect (especially the treatment of gravitation). Secondly, the question, whether or not a given observation is physically possible, is not a question only about the formal apparatus of a theory such as quantum mechanics. It also concerns the nature and arrangement of the physical stuff where the rules and descriptions of theory play out.
In the context of the universe as it is (rather than some other hypothetical universe) it is an open question whether interference experiments of the type under discussion are possible, if `possible' means `physically possible', which in turn means `an apparatus to achieve it could in principle be built'.
In short, it is not known whether or not any amount of technical or technological progress will make it possible to gain empirical evidence of quantum superposition of a cat. More generally,
the approximation of the isolated system may or may not be realisable
in the case of quantum superpositions of large chaotic systems over ordinary distance- and time-scales. 

A related issue has recently been highlighted by Gisin and co-workers. This is that one may reasonably suspect that there
are physical limits on information density. To make this specific and
concrete, they propose what they term the
{\em Principle of finiteness of information density}, which is simply that 
``A finite volume of space can only contain a finite amount of information."
\cite{Gisin2017,Gisin2021}
This is closely related to the Beckenstein bound concerning black holes
in General Relativity. A precise statement is impossible, in our current state 
of knowledge, owing to our lack of a complete understanding of the relationship 
between quantum theory and gravitation. The relevance to the current discussion is that if the principle holds then the Hamiltonian of any system can only
have some finite precision, and in the case of chaotic systems this
implies that many details of the final state cannot be determined from the 
initial state. This would suffice to make it strictly impossible to perform an interference experiment of the
Schr\"{o}dinger cat type (i.e. to obtain empirical evidence that there is a superposition of alive and dead by observing a quantum interference). 

Another noteworthy development is the recent proof, by Bakshi and co-authors,
that thermal states of local Hamiltonians are separable at finite temperature
\cite{Bakshi2024}. To be precise, the authors show that the Gibbs state
$\rho = e^{-\beta H} / {\rm Tr}(e^{-\beta H})$, 
for a local Hamiltonian $H$ on a graph with degree $d$, is a 
classical distribution
over product states for all $\beta < 1/(cd)$, where c is a constant. 
Prior to this work it had been suspected (or widely assumed) that
short-range quantum correlations, like short-range classical correlations, exist at any constant temperature. It was guessed that entanglement would merely become
smaller, or of shorter range, as the temperature of a Gibbs states increases,
but it turns out that the entanglement 
goes strictly to zero at some finite temperature. 
This is connected to the present discussion because it is an
example of entanglement strictly vanishing at a finite value of a relevant
parameter (here, the temperature).

We already remarked, at the start of this section, that quantum 
theory is properly
applied when it is used to describe how parts of the physical 
world influence one another (as opposed to the attempt to describe the
universe as a whole from a perspective outside it). In this regard the following is a noteworthy aspect: even without any modification to existing quantum field
theory, it already predicts information loss whenever it is applied to finite
regions of spacetime \cite{Unruh2017}. Such information loss is not
alien to quantum field theory, but intrinsic to it. In order to highlight this
fact, let's elevate it to a theorem:
\begin{quote}
{\bf Theorem 1.} {\bf QFT information loss}. \em
Information loss is a characteristic feature of ordinary quantum field theory applied 
to finite regions of spacetime. 
\end{quote}
{\bf Proof}: see \cite{Unruh2017}.

Scientific discussion 
in general, and the discussion of quantum theory especially, must take into
account the fact that observations amenable to scientific description
themselves take place within the physical universe. We do not have access
to a perspective from outside the universe, from which the whole universe can
meaningfully be discussed. Any universal physical theory must perforce
be a theory of interacting finite subsystems
which may leak information to their environment. Furthermore, as 
Cuffaro and Hartmann point out, there is no need to assume the evolution of
the cosmos as a whole to be unitary; one can adopt what they term the 
{\em open systems view} and formulate a fundamental theory accordingly
\cite{Cuffaro2024}. The approach arrived at this way is the very one already 
used when we apply quantum theory in practice. In this approach the open system 
description is the accurate one, while the isolated system is an idealization 
and approximation; irreversible behaviour is what happens, reversible behaviour 
is the idealization which offers a good approximation in some circumstances. 

A further property of the universe, namely the cosmic event horizon,  
underlines the fact that some behaviours are irreversible in practice, no matter 
what the equations of motion might be. Not all cosmologies have such a horizon, 
but observational
evidence suggests that the cosmos we inhabit does. It means that, owing
to cosmic expansion, a sufficiently large quantum interferometer cannot
recombine its two `arms' and so an interference effect becomes unobservable.
This is sufficient to render it impossible for the multiple worlds of an
Everettian model to influence one another once the impact of a given process
has extended to a large enough distance. The required distance for this
form of decoupling (a type of decoherence) is
half way to the horizon of the event at which some given
superposition had been brought about.

\subsection{Handling the issue via notation}

I will not, in this paper, assume an answer one way or the other concerning
whether it is physically possible to get empirical evidence of a quantum superposition of large chaotic systems such as some kilograms of water at room temperature, or the body of a mammal. 
But I will adopt a policy of keeping clear, in the
mathematical notation, when processes which are very difficult to control
are being described. This will be done by writing explicitly the phase
$\alpha$ in a superposition such as
\be
( \ket{\phi_1} + e^{i \alpha} \ket{\phi_2} ) / \sqrt{2} ,
\ee
rather than absorbing the phase into the definitions of the terms in the superposition. This policy also offers a way to incorporate the possibility
of non-unitary behaviours if their result is equivalent to randomizing $\alpha$.

\section{Single-world and many-worlds}  \label{s.singlemw}

This section will define the difference between `single-world' and `many-world' quantum theories
and will introduce ideas to be expounded in subsequent sections.

I will discuss the Schr\"odinger cat thought-experiment
using the traditional categories of alive and dead.
Let $A$ and $D$ denote configurations of the matter of a 
cat and its immediate environment which are
consistent with an alive cat and a dead cat, respectively. Thus $A$ 
refers to configurations in which there is ongoing cell chemistry,
a heart beat, circulating blood, etc., and $D$ refers to configurations
where the heart has ceased beating, oxygen levels in the blood are
depleted and cell chemistry rapidly reverts to decay. A central feature of
the \Sch\ cat thought-experiment is the hypothesis that it is in principle
possible to isolate a system such as a cat by placing it in some finite box
whose size and walls are sufficient to prevent entangling interactions
between the cat and other systems outside the box. This hypothesis is questionable (see Section~\ref{s.isolated})
but let us adopt it for the sake of argument. 
Expert opinion is divided on what assertions about the cat
follow from quantum theory. One view is that the evolution of physical stuff
is entirely
unitary, and in consequence both $A$ and $D$ feature in any complete description
of that which exists physically at the end of the experiment (were
the experiment to be performed). This type of interpretation is termed a
{\em relative-state} or {\em many-worlds} interpretation. Another view is that
the physical world evolves in a non-unitary manner and quantum theory describes
this aspect through the way the theory is applied, when it is applied correctly.
On this kind of interpretation, quantum theory warrants statements about what is
so after irreversible processes have taken place, such that a full description
of that which exists physically can include one and only one of $A$ and $D$
at the end of the experiment. This type of interpretation is termed a
{\em one-world} (or {\em single-world}) interpretation.
To the objection that `there are
no strictly irreversible processes', the one-world approach may reply that it
is sufficient if there are processes which are never reversed in fact, and
all human discourse takes place under such an implicit assumption. 

As Deutsch correctly pointed out, the single-world and many-world approaches are
not merely different interpretations of quantum theory but different theories 
because they make different physical predictions \cite{Deutsch1985}.
In order to test empirically which (if any) 
of these theories is correct, one would need to
conduct an experiment which can distinguish the `either/or'
(one-world) outcome from the `both-and' (many-world) outcome. Such a
test would require a further evolution of the cat and its surroundings which 
erases the {\em which-path} (alive or dead) information. For example, if we 
suppose that the mathematical
apparatus of Hilbert-space vectors and unitary evolution applies to cats, then
one might suppose that a state $\ket{\psi_1} = (\ket{A}+e^{i\alpha}\ket{D})/\sqrt{2}$ has
been obtained, but we only have a measuring device which can unequivocally
distinguish $\ket{A}$ from $\ket{D}$. In such a case one might aim to
realise the unitary evolution $\ket{\psi_2} = U \ket{\psi_1}$ with
\be
U = \frac{1}{\sqrt{2}} \left[
\ket{A}\left(\bra{A} + \bra{D}\right) +  
\ket{D}\left(\bra{A} - \bra{D}\right) \right].
\ee
Upon receiving the whole apparatus (nucleus, poison, cat, box)
as it stands at the end of an ordinary \Sch\
cat experiment, one applies to it the operator $U$ (using some highly
sophisticated device) and then measures in the $\ket{A}$, $\ket{D}$ basis. 
If such an experiment were repeated many times (always keeping the relative phase
of $\ket{A}$ of $\ket{D}$ under control) and the set of final measurements
yielded results inconsistent with a fifty percent chance of
finding $A$ then one would infer that the cat's state had indeed been
in superposition before $U$ was performed, as in state $\ket{\psi_1}$.

The difficulty of realising, in the laboratory, an evolution such 
as $U$ was discussed in the previous section. It is also
illustrated by the fact that an apparatus to realise $U$ would
be able to return a dead cat to life, since $U \ket{D} = (\ket{A}
- \ket{D})/\sqrt{2}$. 
We will not further discuss such difficulties here. 
We will discuss how one should reason correctly when
considering experiments of this kind, especially the one which has 
become known as {\em Wigner's friend} \cite{Wigner1967}.

The thought-experiment known as Wigner's friend is
a development from the \Sch\ cat thought-experiment,
designed to bring human observers explicitly into the discussion. 
One supposes that one human being observes the cat and comes to their
own conclusions, and a further human being, who as yet has 
made no observation, wishes to know how quantum theory should be applied 
so as to describe correctly what they can expect to observe. If our
interpretation of the original cat experiment is that one and only one of
$A$ and $D$ takes place, then the experiment of
Wigner's friend introduces no new considerations.
If, on the other hand, one adopts a relative-state interpretation then one
will enter into a discussion designed to show that it is consistent with
human experience. One might, for example, hold that person $W$ can
legitimately
describe person $F$ as being in a quantum superposition involving different
sets of memories, and this is consistent with the fact that 
when consulted, $F$ reports only one of the 
memories (i.e. either `cat lived' or `cat died') in each `branch' or `world'.

I will not rehearse arguments to show that relative-state interpretations
are consistent with human experience. My aim is to show the following:
\begin{enumerate}
\item 
If one assumes that reasoning beings always evolve in a strictly unitary
way (i.e., in accordance with \Sch's equation) then their reasoning
about what the future may hold has to take this into account,
and this implies that statements involving what a given observer
observes must be qualified in a way I shall discuss. 
\item The presence of a permanent record is sufficient to remove the
need for such a qualification. 
\item Such a record also makes it possible to assert a single-world statement of quantum mechanics which does not require modifications to the equation of motion
and which is consistent with relativistic quantum field theory. 
\end{enumerate}
The {\em permanent record} mentioned in item 2 is a lingering change
in the universe which stores {\em which-path} information in a way
I shall discuss. The mention of equation of motion in item 3 draws
a contrast with GRW-like collapse theories. The mention of 
quantum field theory draws a contrast with the 
de-Broglie--Bohm treatment \cite{Wallace2022}.

\section{On permanent records and the projection postulate}  \label{s.project}

The argument of this section will show that the projection postulate that is
invoked in some presentations of the structure of quantum theory can be
clarified by bringing in the notion of a {\em permanent record}.
The projection 
postulate, with its reference to a process called a `measurement' or an
`observation', is, on this argument, a reference to those types of process
which take the generic form we shall describe, in which one thing interacts with
another in such a way that subsequent 
observations are indistinguishable from those which would be obtained if
the projection had occurred.

In order to show this we shall invoke some standard results in the 
discussion of composite quantum systems and decoherence. We 
employ the 
mathematical apparatus of state vector, Hilbert space, and 
Schr\"odinger's equation. It is important to keep 
distinct in ones mind the mathematical abstractions and the physical entities.
The physical entities do whatever they do, by whatever means they do it;
the mathematical abstractions provide a language which can capture the 
behaviour sufficiently for us to make claims which can be empirically 
tested.\footnote{Here I assert a mere truism; it should not be taken to
imply a purely instrumentalist epistemology.} The argument of this section has three main parts.
Part I is purely mathematical and uncontroversial: it is to derive eqn (\ref{rhod}) from (\ref{psin}).
Part II is to argue that, in the case where the process leads to a 
permanent record (to be defined), every observable aspect of the subsequent 
behaviour is consistent with what would be the case if a projection had taken 
place. This ought to be uncontroversial, since it follows from standard 
properties of the density matrix, but an Everettian might wish to claim that
it does not adequately account for other worlds which are unobservable to us but
claimed to be real. 
Part III of the argument is more controversial. I will argue that the above is 
sufficient to warrant a claim that the projection
does in fact take place. However this third part is
not required for the rest of the paper.

{\bf Part I}. Let us consider the case of a composite system $M$ 
made of many two-state systems which we shall call qubits. This will serve
as a model which captures the relevant considerations for a very wide range
of physical systems. (We allow that the qubits may be physically represented by almost anything, including modes of an electromagnetic field, for example, or locations or rotations of massive particles.)
The energy eigenstates of each qubit will be written
$\ket{0}$, $\ket{1}$, and those of a collection of $n$ qubits will be written
$\ket{{\bf x}_n}$ where $\bf x$ is an $n$-bit binary number.
${\bf x} = {\bf 0}$ signifies that all digits of $x$ are $0$;
${\bf x} = {\bf 1}$ signifies that all digits of $x$ are $1$.
Suppose that $M$ is prepared in the state $\ket{{\bf 0}_n}$, and then it
undergoes an interaction with a single further qubit prepared in
$(\ket{0} + \ket{1})/\sqrt{2}$, with the result that the overall state of
both becomes
\be
\ket{\psi(\alpha)} =
\frac{1}{\sqrt{2}} \left( \ket{0} \ket{{\bf 0}_n} + e^{i\alpha}
\ket{1} \ket{{\bf 1}_n} \right).
\label{psin}
\ee
In some circumstances, empirical observations on such a system will be
the same as those which are obtained for a system prepared
in one of the states
$\ket{0}\ket{{\bf 0}_n}$ or $\ket{1}\ket{{\bf 1}_n}$ with equal probability. This
occurs, in particular, when some part of the system is not available to
the one observing. We can model this by supposing that one of the $n$ qubits
has escaped to some location outside the region where a given observer can act,
or perhaps it is there somewhere but it is hard to find and has not been found.
(In practice the lost qubit may be in thermal radiation or a spin precession
or similar). In this case the information that an observer can gather is expressed
by the {\em reduced density matrix}, which is the density matrix obtained
by averaging over the missing information by taking a partial trace. 
After such a trace the density
matrix of the remaining parts (the observed qubit and the $n-1$ other qubits)
takes the form
\be
\rho_d = \frac{1}{2} \left( \begin{array}{cc} 1 & 0 \\ 0 & 1 \end{array}
\right)
\label{rhod}
\ee
where we have written it in the basis $\ket{0}\ket{{\bf 0}_{n-1}}$,
$\ket{1} \ket{{\bf 1}_{n-1}}$.

The density matrix of a system randomly
prepared in just one of
those two states, with equal probability, would take this same form. This
proves that the statistics of all empirical observations on the system (made by the observer 
with no access to the final qubit) are identical to those which would be
made on a system which was randomly prepared in one of the two given
basis states.

A similar analysis applies to the case where an observer has access to all the
qubits, but the phase $\alpha$ is uncontrolled, with the result that it is
a random variable, evenly distributed between $0$ and $2\pi$. The density
matrix for this case is again $\rho_d$, with the basis states now including
all the qubits.

In the above we argued from a particular choice of basis in which to compose the 
density matrix, and this
needs justification. Any density matrix can be expressed as a sum of pure 
density matrices, where a {\em pure} density
matrix is one which can be written in the form $\ket{\psi}\bra{\psi}$ for some
$\ket{\psi}$. The decomposition is not unique. This leads to 
the {\em preferred basis problem} in quantum measurement theory. Numerous analyses
have shown how a preferred basis can be identified in interactions
involving a large composite system. The preferred basis is often called the
{\em pointer basis} \cite{Zurek2003}. In his discussion of the relative-state approach, Deutsch calls
it the {\em interpretation basis} \cite{Deutsch1985}.  It can often be identified as that basis in
which an interaction Hamiltonian is diagonal, or as a set of separable (i.e.
not entangled) states into which the density matrix can be decomposed after
decoherence has taken place \cite{Steane2007C}. 
Here, {\em decoherence} is a generic term for such processes
as the loss of a qubit or the randomization of a phase as described above.

The movement from eqn (\ref{psin}) to eqn (\ref{rhod}) involves the notion
of missing information. One may assert that in truth the system state is
$\ket{\psi(\alpha)}$ and that $\rho_d$ gathers information available to one who 
has incomplete information about the state of the system. 
Their information is incomplete either because they do not have access to
a small part of the system, or because they are unable to predict
or control $\alpha$.

{\bf Part II}.
Now let us consider a further possibility: a {\em permanent record}. Let us 
call the lost or escaped qubit $L$. Suppose that $L$ is {\em never} found, or 
{\em never} returns.
To be precise, suppose that, in the subsequent evolution of the whole physical 
cosmos, either (1) no process ever involves a non-negligible
combined influence from $L$ and 
$M_-$, where $M_-$ is the system $M$ without $L$, or (2) $L$ goes
on to become entangled with further systems and no process 
ever involves a non-negligible combined influence from all those further systems and $M_-$.
In this case no observer will ever be in receipt of information sufficient to
allow them to determine that the state was $\ket{\psi(\alpha)}\bra{\psi(\alpha)}$ and not
$\rho_d$. {\em Therefore the prediction from quantum theory is that future 
observations of the universe will be empirically indistinguishable from those 
obtaining for the case where the state was a
probabilistic mixture (not quantum superposition)
of $\ket{0}\ket{\bf 0}$ and $\ket{1}\ket{\bf 1}$.}

{\bf Part III}.
We now have to deal with one of the central puzzles of quantum measurement theory, which is to show how just one of the two states (chosen randomly) is realised in practice in any one realisation of the measurement. The mere presence of decoherence alone does not provide an answer \cite{Adler2003}. 
Drossel and Ellis express the issue in these terms: ``If the heat bath was described by a many-particle wave function that evolves deterministically in interaction with the electrons, the reduced density matrix of the electrons would
describe a system that is entangled with the environment, and the measurement problem would not be resolved" \cite{Drossel2018}.
The solution proposed by Drossel and Ellis is that quantum theory with unitary dynamics simply does not apply to the type of large physical system under discussion (called by them a heat bath), whereas a stochastic evolution of the type assumed in practice in statistical mechanics does apply,
and we can interpret the density matrix as furnishing the statistics of a set of trajectories, each of which is stochastic \cite{Breuer2002,Gisin1992,Dalibard1992,Tian1992}. This approach has been expounded by Cuffaro and Hartman \cite{Cuffaro2024}, under the title `The Open Systems View'. In this view, the central idea in a correct treatment is not that there is a universal state evolving as $\hat{H} \ket{\Psi}
= i \hbar |\dot{\Psi}\rangle$, but rather, any given physical system is described by a density matrix evolving as
\be
i \hbar \dot{\rho} = [ \hat{H},\;\rho ] + {\cal L} \rho   \label{rhodot}
\ee
where the effect of the superoperator $\cal L$ need not be, and usually is not, unitary. As Cuffaro and Hartman and explain, there is no need to regard (\ref{rhodot}) as imprecise or as inapplicable to the entire universe. (Conservation of probability does not necessarily require unitary evolution, for example.) 

In this paper I am endeavouring to remain agnostic on which of the various
single-world treatments of quantum mechanics should be preferred. My aim is
to show that some thought-experiments which have been claimed to show 
inconsistency in the single-world approach in fact do not do so. Rather than choosing among, for example, Copenhagen, Modal, Transactional, QBism, or Drossel and Ellis,
I merely advocate the following
statement of the way the physical relates to the mathematical in quantum mechanics: 
\begin{quote}
\em Processes leading to a preferred basis and a permanent record are what happens in the world; the mathematical tools leading to the diagonal (or block-diagonal) density matrix are furnishing statements about what basis is preferred, and what is the probability distribution amongst the states of that basis.
\end{quote}
This distinction between physical and mathematical is like, and might be
identical to, the distinction between the {\em value state} and 
{\em dynamical state} in modal interpretations of quantum mechanics. 
Alternatively, the above statement can be read as an axiom of the Copenhagen
Interpretation, or of a modern Copenhagen-like interpretation. My claim is that any single-world interpretation will either adopt some such statement as the above, or will propose a process whose outcome is empirically equivalent to the above, and the resulting theory is entirely in agreement with what is observed. 

On an open systems view, physical evolution is a stochastic process and quantum 
mechanics describes the evolution of the density matrix via (\ref{rhodot}). But no matter what mathematical method is used, the statement of the theory has to assert how the mathematical quantity (the density matrix) relates to the physical situation. A standard practice is to assert that the density matrix gives the probability distribution for a conceptual ensemble (a Gibbs ensemble) of copies of the physical system in question. In this case 
the replacement of $\rho_d$ by one of $\ket{0}\ket{\bf 0}$ 
and $\ket{1}\ket{\bf 1}$ (in any one realisation of the experiment) is not a physical process. It is a housekeeping step
in the mathematical analysis. It just tidies up the notation. Peres \cite{Peres1986}, for example, expounds this point. 

The terminology `state vector' or `state', in reference to a Dirac ket (a vector or ray in Hilbert space) is somewhat misleading.
As Peres put it: ``This use of the word ``state" is unfortunate, as it prejudged the actual meaning of the Hilbert space rays." \cite{Peres1984sv,Peres1986}.
The ket is not the state, but rather a way to find out what are the possible states. A ket such as $\ket{\psi(\alpha)}$
does not necessarily endorse a statement that the physical cosmos is in
superposition. This is because its physical interpretation can only be established
with reference to whatever permanent records are being brought about. 
The physical state is not a vector in Hilbert space; it is a distribution of
worldlines in spacetime. The vector in Hilbert space is part of a mathematical apparatus which furnishes the probability distribution among those outcomes
which are never erased.

To make the same point another way: there is no particular time at which one need assert 
that a collapse takes place (evolving $\ket{\psi(\alpha)}$ to one of $\ket{0}\ket{\bf 0}$ or
$\ket{1}\ket{\bf 1}$). It suffices to note that if all empirical observations forever
will be correctly predicted by one of $\ket{0}\ket{\bf 0}$ or $\ket{1}\ket{\bf 1}$ 
holding at some time $t$ then we
have warrant to say things like `this is the outcome that happened'. The famous question,
`where does the chain of measuring apparatuses (the von Neumann chain) stop?' is like the question `where
precisely is the black hole horizon?' The answer to the second question involves what
happens to the black hole into its infinite future, but this does not render the theory
of general relativity either implausible or impossible to apply. The von Neumann chain
(which in practice is a chain of ordinary, not engineered, physical processes) is
similar.

Deutsch has suggested that, in order to empirically verify the relative-state 
(many-world) theory it would be sufficient if computer technology advanced to 
the point where we would have warrant to regard controllable machines such as 
quantum computers as the kind of entities which are
commonly called `observers' \cite{Deutsch1985}.
Then one replaces the cat in the \Sch\ cat experiment by such
a machine, and one would have unitary evolution in which, he argues, all the 
relevant aspects of the physics are on show. However the latter is the very 
thing which is in question. {\em The definite
outcome envisaged in a one-world approach occurs precisely in those 
circumstances where
physical evolution is not so well-controlled as to be reversible.}

\subsection{Quantum erasure}

Quantum erasure will feature largely in the rest of the paper, so we shall define it. 

Consider again the state $\ket{\psi(\alpha)}$ defined in eqn (\ref{psin}). It will be useful to extend the notation by attaching subscripts to indicate the two interacting systems:
\be
\ket{\psi(\alpha)} =
\frac{1}{\sqrt{2}} \left( \ket{0}_S \ket{{\bf 0}_n}_M + e^{i\alpha}
\ket{1}_S \ket{{\bf 1}_n}_M \right).                 \label{SMentangle}
\ee
The concept of quantum erasure turns on the following identity:
\be
\ket{\psi(\alpha)} \equiv
\frac{1}{2}\left[\left(\ket{0}_S + e^{i\alpha} \ket{1}_S \right)
\ket{\pmb{+}_n}_M  +
 \left(\ket{0}_S - e^{i\alpha} \ket{1}_S \right)
\ket{\pmb{-}_n}_M \right]         \label{qerase}
\ee
where
\be
\ket{\pmb{+}_n}_M \equiv \frac{1}{\sqrt{2}} \left( 
\ket{{\bf 0}_n}_M  + \ket{{\bf 1}_n}_M \right),
\qquad
\ket{\pmb{-}_n}_M \equiv \frac{1}{\sqrt{2}} \left( 
\ket{{\bf 0}_n}_M  - \ket{{\bf 1}_n}_M \right).
\ee
An apparatus that has access to $S$ and all of $M$ could reveal 
information about the value of
$\alpha$ (e.g., it could distinguish the case $\alpha = 0$ from the case
$\alpha = \pi$). 
One way to do this is to bring about a unitary evolution
which disentangles $S$ from $M$. But, in view of (\ref{qerase}), another
way would be to perform a measurement of $M$ in the $\{\ket{+_n},\,
\ket{-_n}\}$ basis. On a projective analysis of measurement, the result will
be one of
\be
\frac{1}{\sqrt{2}} \left(\ket{0}_S + e^{i\alpha} \ket{1}_S \right)
\ket{\pmb{+}_n}_M  \qquad \mbox{or} \qquad
\frac{1}{\sqrt{2}} \left(\ket{0}_S - e^{i\alpha} \ket{1}_S \right)
\ket{\pmb{-}_n}_M \, . 
\label{erase}
\ee
This way of proceeding is commonly called
{\em quantum erasure}. The idea is that the which-path information which
was contained in the state of $M$ has been erased, and consequently
the state of $S$ is a superposition in which the relative phase $\alpha$
can influence subsequent observations of $S$, for any observer who
knows which of $\ket{\pmb{+}_n}_M$ or $\ket{\pmb{-}_n}_M$ was obtained.

It will be important, in the following, to note that quantum erasure
completely removes from $M$ any lingering impact of its interaction with
$S$. At the end no part of $M$ stores any indication of any influence 
$S$ may have had. In the rest of the paper we shall use the terms 
`erasure' and
`quantum erased' for any evolution which has this property. It will not
be necessary to distinguish the cases where the erasure was brought about
by a unitary evolution or by a projection. 

Equations (\ref{SMentangle})--(\ref{erase}) serve to {\em define}
quantum erasure. They make no claim about whether such erasure is possible
in a given case. It is highly significant that in the presence of a permanent
record of the type discussed above, such erasure is impossible. We already
noted that a process leading to a permanent record cannot be reversed
(by definition); we now note, further, that its outcome 
cannot be quantum-erased either. This is because such erasure requires both
complete control over M and a stable, reproducible $\alpha$, either or
both of which is not the case when there is a permanent record.

\section{The extended Wigner's friend thought experiment}  \label{s.extendW}

In the remainder of the paper we show how the single-world quantum theory
described above can be applied to the extended Wigner's friend thought 
experiment of FR, and similar thought experiments.

Frauchiger and Renner have proposed a thought experiment which
takes the form of an extension to the Wigner's friend scenario. Their
argument is concerned with reasoning on the part of different observers
in a specific set of events and it aims to show that, for these events,
(quoting from the abstract), 
``one agent, upon observing a particular measurement outcome, must conclude that another agent has predicted the opposite outcome with certainty. The agents' conclusions, although all derived within quantum theory, are thus inconsistent. This indicates that quantum theory cannot be extrapolated to complex systems, at least not in a straightforward manner."\cite{Frauchiger2018}
In this and subsequent sections I will re-analyse this 
thought-experiment and use it to motivate argument about what it may mean to
be an observer who can reason and who can communicate their experiences
to other interested observers.

FR claimed on the basis of their analysis that single-world interpretations 
of quantum theory cannot avoid inconsistency if they retain the standard
quantum mechanical apparatus of \Sch\ equation and Born rule. 
Sudbery then pointed out that this cannot be the right way to state the
conclusion, since the theory formulated by Bohm and elaborated by Bell 
shows that 
{\em any quantum system can be modelled in such a way that there is only one 
``world" at any time, but the predictions of quantum theory are 
reproduced} \cite{Sudbery2017,Lazarovici2019}.
This does not deny that FR's thought-experiment has useful lessons to offer. 
FR's argument works by showing that a certain set of ideas is self-contradictory. I will show that
the particular set they alluded to is not self-contradictory if one attaches
the notion of a single world to the notion of a permanent record. 

Kastner \cite{Kastner2020} argues that the problem lies in the claim
that physical evolution is unitary. That is, the conclusion should be 
precisely the opposite of the one recommended by FR: it is the entirely unitary Everettian picture which does not make sense.  
In the present discussion I will show what happens when the processes inside the laboratories are unitary, and what happens when they are not. I don't require a specific model of the non-unitarity; it is sufficient if its effects are equivalent to those of a projection onto a preferred basis (pointer basis). Kastner also points out, {\em contra} Sudbery,  that it is not so clear whether de-Broglie Bohm can give a consistent account.

The three conditions proposed by FR can be expressed as follows (here I adopt
my own words which are based on a combination of those in FR and those in
Sudbery's formulation)
for a given theory~$T$:
\begin{itemize}
	\item {\bf QT} {\em Compliance with quantum theory}: In $T$,
\Sch's equation stands, unmodified by spontaneous or gravity-induced 
collapse processes and the like, and if an agent knows that a 
system's state is an eigenstate of some observable, then they can also know that
a measurement would yield the corresponding eigenvalue.
	\item {\bf SW} {\em Single world:} 
$T$ denies that measurement interactions lead to multiple
outcomes in the way affirmed by relative-state or many-world interpretations.	
	\item {\bf SC} {\em Self-consistency}: $T$?s statements about measurement 
	outcomes are logically consistent (even if they are obtained by considering the 
	perspectives of different experimenters). [For example, suppose that agent $A$ has established that
``I am certain that agent $B$, upon reasoning within $T$, 
is certain that $x = \xi$ at time $t$" then agent $A$ can conclude that
	``I am certain that $x = \xi$ at time $t$."]
\end{itemize}
FR's argument that these are mutually contradictory involves an implicit assumption
that the physical processes involved in the thought-experiment could all in principle
be reversed or undone by quantum-erasure. 
However, this ought not to be assumed, but rather treated on a case-by-case basis. 

Healey presents the FR thought-experiment in such a way as to avoid one
aspect of their presentation, which enables him to clarify the reasoning process \cite{Healey2018}.
In his discussion, Healey prefers not to adopt the 
above formulation in terms of QT, SW, and SC, and instead invokes the following:
\begin{itemize}
\item {\bf Universality} Quantum theory may be applied to all systems, including macroscopic
apparatus, observers and laboratories.
\item {\bf No collapse} When an observable is measured on a quantum system in a physically
isolated laboratory, the state vector correctly assigned by an external observer to the
combined system+laboratory evolves unitarily throughout.
\item {\bf Unique outcome} A measurement of an observable has a unique, physical outcome.
[But see remarks below]
\end{itemize}
The 2nd item, {\em no collapse}, here means precisely that the outcome of a process
termed `measurement' is treated by an entangled state such as (\ref{SMentangle}), but
note that Healey is careful to say `the state vector correctly assigned by an external observer to'
and not simply `the state vector of'. One might suspect that this cannot be consistent
with the 3rd item, {\em unique outcome}, but the `unique outcome' affirmed by Healey is
{\bf not} the same as SW. `Unique outcome' in his sense means merely that every
`branch' in a set of multiple universes has a copy of an agent finding that they observe one 
outcome not many (but the outcomes may differ from one branch to another).
In other words the role
of `Unique outcome' is to ensure consistency with human experience, not consistency with
a single-world approach to quantum theory. In view of this, Healey does not reproduce 
FR's argument but instead sets out another way in which reasoning from quantum theory
can go wrong.

Healey's contribution clarifies certain issues. In particular he emphasizes the care that
is needed when arguing from counter-factuals concerning quantum systems and their
correlations. An argument taking the form `if $A$ had been the case then $B$ would have been 
the case' may be invalid if $A$ involves a system which is entangled with some third
party and that entanglement has not been correctly incorporated into the argument.

Baumann, Hansen and Wolf discuss further aspects \cite{Baumann2016}.
They take the view that 
a consistent application of a single-world interpretation will
not take the form suggested by FR.
FR describe what amounts to wavefunction
collapse (by arguing that, if SW is upheld, then an observer $F$ must observe one and only one
outcome of a measurement), but then suggest that a second observer $W$
will consider that $F$ and the environment of $F$ together evolve
unitarily. This idea (i.e. the notion that collapse of a quantum state
is a subjective rather than objective matter)
is called by Baumann {\em et al.} {\em subjective 
collapse} and they argue that it is here that the inconsistency arises.
On this view, the FR thought-experiment serves to show that
subjective collapse cannot consistently be invoked in discussions of
quantum theory. I agree with that conclusion.

Baumann {\em et al.} argue that an analysis 
involving projection upon observation reaches no contradiction, but 
makes a different prediction for 
the physical behaviour than the one obtained by assuming unitary evolution
throughout. This is correct.
A naive reaction would be to infer that 
only one of those two versions of quantum theory can be right. 
But we already saw
in Section~\ref{s.project}
that unitary evolution can (and often does) give rise
to outcomes which are empirically indistinguishable from evolution with 
projection.  Therefore an analysis which invokes projection
can be correct when those circumstances hold, and not when they do not.

\section{The protocol and the argument of Frauchiger and Renner} \label{s.FR}

We shall now describe the FR thought-experiment
in notation which is intended to be as succinct and clear as 
possible while retaining the salient points.

\begin{figure}
	\myfig{0.3}{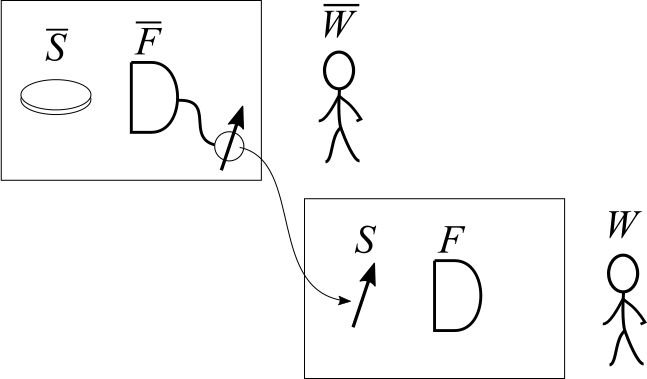}
	\caption{Schematic diagram of the systems envisaged in the argument in the text. $\overline{S}$
	is a (quantum) coin; $S$ is a spin-half which starts in the upper laboratory and is subsequently
brought into the lower laboratory. $\overline{W}$ and $W$ are people (e.g. human beings); $\overline{F}$ and $F$
are fully controllable quantum systems in the first treatment, eqns (\ref{step1})--(\ref{step5}),
and then they are replaced by human beings in the second treatment, eqn (\ref{psi5tilde}).}
	\label{f.boxes}
\end{figure}

Suppose that we have a quantum system which can be divided into 
four parts, labelled $\overline{F}$, $\overline{S}$,
$F$, $S$ (see figure \ref{f.boxes}). The parts $\overline{S}$ and $S$ are two-state systems. 
States $\{\ket{h},\;\ket{t} \}$ form an orthonormal basis for $\overline{S}$. States $\{ \ket{\up},\;\ket{\dn} \}$ 
form an orthonormal basis for $S$. (You can, if you
like, consider $\overline{S}$ to be a simple `coin' and its states are
`heads' and `tails'). Systems $\overline{F}$ and $F$ each have a state $\ket{i}$ (for `init' or `initial') and also further states.

In what follows, $\overline{F}$ will interact with $\overline{S}$ and $F$ will interact
with $S$ in a simple kind of unitary process called a {\em measurement-like} 
interaction, which is very similar to the {\em controlled-not} logic gate
in quantum computing. For our purposes these unitary processes
$\overline{U}$ and $U$ are sufficiently 
described by listing what happens in the following cases:
\be
\overline{U} \ket{i}_{\overline{F}} \ket{h}_{\overline{S}} &=& 
 \ket{\vh }_{\overline{F}} \ket{h}_{\overline{S}}, \;\;\;\; \nonumber
\overline{U} \ket{i}_{\overline{F}} \ket{t}_{\overline{S}} \;=\; 
\ket{\vt }_{\overline{F}} \ket{t}_{\overline{S}};  \;\; \label{Ubar} \\
U \ket{i}_{F} \ket{\up}_{S} &=& \ket{\Up}_{F} \ket{\up}_{S}, \;\;\;\;
U \ket{i}_{F} \ket{\dn}_{S} \;=\;  \ket{\Dn}_{F} \ket{\dn}_{S}  . \;\; \label{Umeas}
\ee
Thus the interaction $\overline{U}$ `copies' the basis states of $\overline{S}$ onto
$\overline{F}$, and similarly for $U,\;S,\;F$.

Systems $\overline{F}$ and $\overline{S}$ together constitute `lab $\overline{L}$';
systems $F$ and $S$ together constitute `lab $L$', but note: system
$S$ begins in lab $\overline{L}$ where it interacts with $\overline{F}$, after which
it is moved to lab $L$ where it subsequently remains. In order to track
the position of $S$ in the notation, we shall adopt the following shorthand:
\begin{eqnarray*}
\ket{\mbox{spin }\up} \otimes \ket{\mbox{position near to $\overline{F}$}} &\equiv& \ket{\up'}_S \\
\ket{\mbox{spin }\dn} \otimes \ket{\mbox{position near to $\overline{F}$}} &\equiv& \ket{\dn'}_S \\
\ket{\mbox{spin }\up} \otimes \ket{\mbox{position near to $F$}} &\equiv& \ket{\up}_S \\
\ket{\mbox{spin }\dn} \otimes \ket{\mbox{position near to $F$}} &\equiv& \ket{\dn}_{S} 
\end{eqnarray*}

In order to write the state of the complete system of four parts, we use
the notation:
$$
\ket{\Psi\, \psi, \, \Phi\, \phi} \equiv
\ket{\Psi}_{\overline{F}} \otimes \ket{\psi}_{\overline{S}}
\otimes \ket{\Phi}_{F} \otimes \ket{\phi}_{S}
$$
(This notation is reminiscent of the practice in quantum information theory of
writing lists of binary digits in order to indicate the state of sets of qubits.)

We now consider a sequence of steps. We will specify the steps precisely in
a moment, but it may be helpful to have a rough statement first. The sequence is:
``prepare coin; observe coin; prepare spin accordingly; move spin to other lab;
measure spin". All the terms in this rough statement should be understood to
be being used loosely. The more precise statement is as follows.

Step 1: ``prepare coin". The system $\overline{S}$ (the `coin') is prepared in a superposition of heads and
tails, with unequal weights; the spin $S$ is down and located near to $\overline{F}$, 
and the other systems are initialized. The resulting initial state is
\be
\ket{\psi_1} = \sthird \ket{i\,h,\,i\,\dn'} + \stthird \ket{i\,t,\, i\, \dn'}.   \label{step1}
\ee
Step 2: ``observe coin". $\overline{F}$ undergoes a measurement-like interaction with the `coin' $\overline{S}$,
with the result that the overall state evolves to
$$
\ket{\psi_2} = \sthird \ket{{\bf h}\,h,\,i\,\dn'} + \stthird \ket{{\bf t}\,t,\, i\, \dn'}.
$$
Step 3: ``prepare spin accordingly". 
$\overline{F}$ interacts with the spin $S$. The interaction is one which
leaves the state $\ket{\vh,\, \dn'}$ unchanged, and evolves the 
state $\ket{\vt,\, \dn'}$ to 
 $\ket{\vt,\, \rightarrow'} \equiv (\ket{\vt,\, \dn'} + \ket{\vt,\, \up'} )
 / \sqrt{2}$. Hence the outcome for the composite system is that the overall
state evolves to
$$
\ket{\psi_3} = \sthird \! \ket{\vh\,h,\,i\,\dn'} + \sthird \! \left(
\ket{\vt\,t,\, i\, \dn'} + \ket{\vt\,t,\, i\, \up'} \right). \;\;
$$
Step 4: ``move spin to other lab". The position of the spin $S$ is moved from
near to $\overline{F}$ to near to $F$. The state evolves to
$$
\ket{\psi_4} = \sthird \left[\rule{0pt}{2.3ex} \ket{\vh\,h,\,i\,\dn} + 
\ket{\vt\,t,\, i\, \dn} + \ket{\vt\,t,\, i\, \up} \right].
$$
Step 5: ``measure spin". Systems $F$ and $S$ undergo a measurement-like interaction;
the state evolves to
\be
\ket{\psi_5} = \sthird \left[\rule{0pt}{2.3ex} \ket{\vh\,h,\,\Dn\,\dn} + 
\ket{\vt\,t,\, \Dn\, \dn} + \ket{\vt\,t,\, \Up\, \up} \right].
\ee
This is equivalent to eqn (12) in Healey \cite{Healey2018}.

We now suppose that two observers make measurements on this composite system.
Observer $\overline{W}$ measures $\overline{L}$; observer $W$ measures $L$. The measurement
basis adopted by $\overline{W}$ is one which includes the following states:
$$
\ket{+'} \equiv \stwo \left( \ket{\vh h} + \ket{\vt t} \right), \;\;\;
\ket{-'} \;\equiv\; \stwo \left( \ket{\vh h} - \ket{\vt t} \right).
$$
The measurement basis adopted by $W$ is one which includes the following states:
$$
\ket{+} \equiv \stwo \left( \ket{\Up\up} + \ket{\Dn\dn} \right), \;\;\;
\ket{-} \;\equiv\; \stwo \left( \ket{\Up\up} - \ket{\Dn\dn} \right).
$$
We have
\be
\ket{\psi_5} = \sqf{6} \left(\rule{0pt}{2ex} 2\ket{+',\,\Dn\dn} 
+ \ket{+',\,\Up\up} - \ket{-',\,\Up\up}   \right)              \label{step5}
\ee
therefore the outcome of the measurement carried out by $\overline{W}$ is

\begin{tabular}{lll}
either: & 
obtain $\sqf{5} \left(\rule{0pt}{2ex} 2\ket{+',\,\Dn\dn} 
+ \ket{+',\,\Up\up} \right)$ & \\
& (with probability $5/6$)  \\
or: &
obtain $\ket{-',\,\Up\up}\;\;$ (with probability $1/6$) .
\end{tabular}\\
We shall say that the value of the measured observable is {\em plus} in the
first case, and {\em minus} in the second. (In \cite{Frauchiger2018} these outcomes were
named `fail' and `ok' respectively.) 

In the case where {\em minus} was obtained by $\overline{W}$, the measurement
performed by $W$ on lab $L$ will yield {\em plus} or {\em minus} with
equal probability. Hence the overall probability for the observations of
$\overline{W},W$ to be \{{\em minus, minus}\} is $1/12$.

So far the discussion has concerned simple quantum systems which could
in principle be fully controlled in an experiment. The systems $\overline{F}$ and
$F$ could, for example, be spin-1 particles, and the systems $\overline{S}$ and
$S$ could be spin-half particles. (Indeed, one could propose an almost
identical protocol using spin-half particles throughout; nothing in the argument
will require that the initial state $\ket{i}$ be orthogonal to
the other states.) The above is standard (and simple) quantum theory; the
controlled evolution of a system from $\ket{\psi_1}$ to $\ket{\psi_5}$,
and its subsequent measurement, is within the competence of several modern
quantum computing laboratories, and there is no controversy over what
will be observed nor on the correctness of the above calculation of the 
outcomes of the measurements by $W$ and  $\overline{W}$. We have not
yet introduced the scenario proposed by FR.

We now turn to the scenario proposed by FR.
It concerns a much more complicated and difficult
experiment. We propose now
that systems $\overline{F}$ and $F$ are rational observers,
that is, physical entities which can acquire knowledge, form rational judgements
and arrive at well-motivated beliefs by means of reasoned arguments. 
Instead of the measurement-like interactions $\overline{U},\,U$
(eqn (\ref{Umeas}))
which appeared in steps 2 and 5 respectively of the protocol, 
we now propose that there
are measurement interactions: that is, an evolution in which system $\overline{F}$
(now called Fionabar) acquires knowledge of the state of the coin, and one
in which system $F$ (now called Fiona) acquires knowledge of the state of the
spin. In a Copenhagen-like interpretation, such measurements could be 
modelled mathematically by establishing a measurement basis at the outset,
and interpreting the task as one of finding probabilities for outcomes
in that basis; the measurement outcomes then form the starting states for
subsequent processes. In an Everett-like
interpretation, measurement processes are modelled by unitary operators but one 
should keep in the mathematical description some allowance 
for the complexity of the complete process. This can be done by supposing
that the evolution is like that given by $\overline{U},\,U$, but introduces
a phase which is very hard (perhaps impossible) to control or predict. The resulting
unitary evolution is described by unitary operators $\overline{M},\,M$:
\begin{eqnarray*}
\overline{M} \ket{i}_{\overline{F}} \ket{h}_{\overline{S}} &=& 
 \ket{\vh}_{\overline{F}} \ket{h}_{\overline{S}}, \;\;\;\;
\overline{M} \ket{i}_{\overline{F}} \ket{t}_{\overline{S}} \;=\; e^{i \overline{\alpha}}
\ket{\vt}_{\overline{F}} \ket{t}_{\overline{S}};   \label{Mbar} \\
M \ket{i}_{F} \ket{\up}_{S} &=& \ket{\Up}_{F} \ket{\up}_{S}, \;\;\;\;
M \ket{i}_{F} \ket{\dn}_{S} \;=\; e^{i \alpha}
 \ket{\Dn}_{F} \ket{\dn}_{S}  ,  \label{Mmeas}
\end{eqnarray*}
where $\overline{\alpha}$ and $\alpha$ are phases which are liable to be randomized
in practice, owing to uncontrolled effects in the experiment. In this case
the state at the end of step 5 of the protocol will be
\be
\ket{\tilde{\psi}_5} = 
\sthird \left[\rule{0pt}{2.3ex} e^{i\alpha} \ket{\vh\,h,\,\Dn\,\dn}  
+ e^{i(\overline{\alpha} + \alpha)}\ket{\vt\,t,\, \Dn\, \dn} 
+ e^{i\overline{\alpha}} \ket{\vt\,t,\, \Up\, \up} \right].                \label{psi5tilde}
\ee

We now introduce the argument of Frauchiger and Renner, which leads to the
contradiction which is the heart of their result. In order to make
the argument, it is necessary to assume that $\overline{F}$ and $F$ can
be reasoning beings (Fionabar and Fiona) while simultaneously assuming
that their physical states can be expressed by state vectors in a Hilbert
space and that
the phases $\overline{\alpha}$ and $\alpha$ could be controlled or predicted.
I do not necessarily endorse or accept those assumptions; my immediate
purpose is merely to show what follows if one does accept them,
according to Frauchiger and Renner. That is, my immediate purpose is
to present their argument, as it would be stated in the terms adopted
in the present paper. To this end,
let us assume that the phases have the value
$\overline{\alpha} = \alpha = 0$. We assume that all four of $\overline{F},\,F,\overline{W}$
and $W$ know the full protocol which they have agreed beforehand, and we
adopt the names Wilbur and Wilburbar for $W$ and $\overline{W}$.
Under these assumptions, when Wilburbar ($\overline{W}$) performs
his measurement, if he obtains {\em minus} then the state observed
by him is $\ket{-',\, \Up\up}$ so he might reason as follows.
\begin{quote}
``I, Wilburbar, observe {\em minus}, and I know what the protocol was,
so I am confident that the state of the two labs is now
$\ket{-',\, \Up\up}$. In particular, I know that the situation in
the other lab is that the spin $S$ is up, and furthermore Fiona 
has observed it to be up and can know this and reason accordingly. 
So she will reason as follows:
\begin{quote} 
	(Argument A1):
	``I, Fiona, observe the spin to be up. But since I
know the protocol, I know that I can only get a spin up in the case
where Fionabar observed tails and prepared the spin accordingly. 
So Fionabar considers that the state she passed to me was not
$\ket{\psi_4}$ but rather
\be
\ket{\psi_4'} = \stwo \left( \ket{\vt t,\, i \,\dn} + \ket{\vt t,\,i\,\up}\right).
\ee
Therefore, if she thinks that my evolution is unitary evolution described by quantum theory with phase $\alpha=0$,
she will form the conclusion that after my interaction with the spin the 
joint state will be
\be
\ket{\psi_5'} = \stwo \left( \ket{\vt t,\, \Dn\dn} + \ket{\vt t,\Up\up}\right)
\;=\; \ket{\vt t,\,+}.
\ee
Fionabar will therefore conclude that when Wilbur comes to measure my
lab, he is guaranteed to observe {\em plus}. But if Fionabar, whom I know
to be rational, forms a conclusion with certainty, then I too must form that
same conclusion (if reason is to be respected), so I too become certain that
Wilber will obtain the outcome {\em plus}."
\end{quote}
Since I, Wilburbar, know all this, I now know that Wilbur will observe {\em plus}."
\end{quote}
We can suppose further that Wilburbar announces his measurement result to Wilbur
before Wilbur carries out any measurement. Wilburbar could also announce the
above reasoning, or Wilbur could figure it out for himself. In either case,
when Wilburbar observes {\em minus}, Wilbur becomes convinced that he must,
with certainty, find his measurement outcome will be {\em plus}. But we already
showed that there is an overall probability of $1/12$ that the 
measurements performed by $\overline{W},W$ will yield {\em minus, minus}. Therefore
if the protocol is repeated many times, eventually a contradiction will
arise: the above argument causes Wilbur confidently to expect
{\em plus}, and yet he observes {\em minus}.

According to the argument in which the evolution is unitary up until
Wilburbar acts, Fiona and $S$ are considered by Wilburbar to be
in the state $\ket{\Up\up}$ after he completed his measurement (and obtained
{\em minus}), and according to the argument in which Fionabar observes tails,
Fiona and $S$ are considered by Fionabar to be in the state $\ket{+}$ 
before the measurement by Wilburbar. The lines of argument leading to
these beliefs are not mutually
consistent. For Wilburbar's measurement of lab $\overline{L}$ can only
lead to his conclusion about the state of lab $L$ if
the labs are in an entangled state before Wilburbar acts.
But if Fionabar's initial observation of the coin leads to only
one outcome, either heads or tails, then the labs will not 
subsequently be entangled. This illustrates that when there is a contradiction, the inconsistency can be located in more than one way.

FR proposed that the conditions QT, SW, SC   
lead to reasoning such as that ascribed to Wilburbar above, and
hence to a contradiction. I agree that the argument we have 
ascribed to Wilburbar is faulty. However I will show that
the fault can be avoided while
still retaining all of QT, SW, SC, correctly understood.
I will argue that when we invoke a concept such
as `observe', and discuss reasoning to conclusions based on observations,
we need to be explicit about whether the observations leave permanent
records. If there is no permanent record then the reasoning process has
to proceed differently than if there were a permanent record.

\subsection{Correct logic in conditions of unitary evolution}  \label{s.Ev}

In this section I shall present an argument
in which the entities we have named Fionabar and Fiona can evolve in accordance
with a sequence of logical steps, but I will not insist that this sequence
amounts to a form of reasoning, or the capacity to understand. It will be sufficient
to our purpose if $\overline{F}$ and $F$ are automated logical processing devices,
or symbolic manipulation machines. I leave it
open whether or not such things can attain mathematical and logical insight, 
and I will leave it open whether or not human beings are such things.

The roles of $\overline{F}$ and $F$ could
be played by quantum computers. These are devices whose evolution is assumed,
for present purposes, to
be unitary, with all relevant interference phases under experimental control.
In this case, by assumption, the evolution is as described in steps 1--5 
listed above, and the state before measurements by Wilburbar and Wilbur is
$\ket{\psi_5}$ not $\ket{\tilde{\psi}_5}$. We can suppose that these quantum
computers are themselves programmed with a sequence of logical operations
that amounts to a slavish form of reasoning. As I have said, we need
not enter into the subtleties of
consciousness here, nor into the question of whether an algorithmic
device is capable of genuinely rational thought. All we need agree is that it
is in principle possible to programme a quantum computer much as one might
program a chess computer: it could be assigned an algorithm whereby various
actions result from various inputs. For example a quantum computer could be
programmed to cause a loud-speaker to blurt out the statement `Wilbur will
observe {\em plus}' or `Wilbur will observe {\em minus}' or `I can't tell for 
sure what Wilbur will observe.' But in the protocol we have described, any
such statement will itself have to be reversible by the measurement carried out
by Wilburbar.

As a reminder that we are now considering that $F$ and $\overline{F}$ might
be quantum computers, I will avoid the names Fionabar and Fiona, but
it will be helpful to retain the use of the pronoun `she'. Now phrases such as
``I see $x$" and ``therefore" are no more than a convenient shorthand 
(perhaps an overly colourful one) for the 
evolution of the internal states of these entities. We will take it
that the entities have been constructed so as to evolve internally
according to a program which causes their internal states to map a sequence
of logical steps. That the steps are indeed logical is affirmed by whoever
provided the program. One might call these machine-like entities
`robots' or `daemons'. It is not necessary to assume that
they have conscious experience; for this reason we will not call them
`observers'. 

Robots of this type can be in superpositions of the states of any basis.

When the evolution of a robot, as defined, invokes a subroutine 
prompted by the 
condition `I observe $x$', the most that can be said is not that
the joint state of robot and observed system is $\ket{\Psi_x} \otimes \ket{x}$
but rather that the joint state is an entangled state in which this is one of
the terms. In this case robot $\overline{F}$ might argue, in the evolution of that 
part of the total state in which she sees tails,
\begin{quote}
	(Argument A2):
``I see tails, but my symbolic processing takes into account
the subtleties of quantum theory and accordingly I reason
(that is, the internal state of $\overline{F}$ so develops)
that this part of my overall state 
is consistent with the claim that Wilbur's experiences will be correctly 
predicted by quantum
theory using the state $\ket{\psi_2}$ at step~2. Therefore I do not
infer that the situation at step 4 is $\ket{\psi'_4}$; I
accept that it is $\ket{\psi_4}$, or at least that this is the right
state vector with which to predict outcomes of Wilbur's measurements. 
Therefore I cannot make any statement with certainty about what Wilbur will 
find. He might observe plus, and he might observe minus." 
\end{quote}
A robot $\overline{F}$ programmed to follow correct logic must reason in this way, 
under the assumptions of unitary evolution and controlled phase. According to 
this argument (argument A2) the internal state configuration corresponding
to ``I see tails" is not sufficient
grounds for Fionabar to infer that the state becomes $\ket{\psi'_4}$
at step 4. This is the point that may strike one as surprising. 

In order to check the overall
consistency we must ask whether $\overline{F}$'s memory of having seen tails
survives into the future when such an experiment is done. Wilburbar's
measurement outcomes are drawn from the basis $\ket{\pmb{\pm}'}$. In either
basis state both heads and tails occur, and the
associated states $\ket{\vh}$ and $\ket{\vt}$ must include every
aspect of $\overline{F}$'s physical state, including her memory. 
A suitably arranged subsequent observation of either $\overline{F}$ or the coin
could then find
them to be in either one of the states $\ket{\vh h}$ and $\ket{\vt t}$.
If this were to happen then $\overline{F}$ in her new
situation will be able to make sense of her memories by
realising that Wilburbar's measurement system was sufficiently invasive
as to be capable of perfectly erasing and fabricating memories for her.

In this physical argument, one must be careful not to imply the presence
of a further layer of reality, in addition to the one described by the state
vectors, in which something like consciousness resides. Rather, a statement
such as ``I see tails" is commentary which a physical system makes
by adjusting other physical systems (e.g. by making noises or marks
on paper, or just by firing neurons), and such commentary can be part of a
large entangled state,
in which some other part may very well be reporting ``I see heads".
This second report will also be made by adjusting physical apparatus such as
air, paper or neurons. Any experimental outcome which depends on an
interference between two such reports must bring all the 
physical systems together so that there is no remaining `which path'
information. This could in principle be achieved if systems $F$ and
$\overline{F}$ were in fact quantum computers.
If such experiments were to be done with human beings
(if we assume for a moment that human memory can be in superposition),
then the conscious experience, with its memories,
is itself being manipulated.

It might be supposed that this use of quantum computers will allow us to settle
the question of how quantum mechanics is best interpreted. It
is not so. For, the behaviour of the quantum computers will depend on how they
are programmed. Will they be programmed to `reason' (i.e. automatically follow
the line of argument) according to argument A1 or argument A2? If they are 
programmed in
advance by human programmers, then they will follow whichever line those human
programmers choose to provide. If they are programmed by training, then the
argument to which they eventually give most weight will be the one which best
enables them to meet whatever goals were set up by their trainers. The goal
of avoiding the `check-mate' condition of inconsistency is here achieved
by adopting argument A2 (or an equivalent such as A3 described below). 
In this case
the quantum computers will give programmed responses that are like those
which a rational agent would make if that rational agent took into account
the conditions of the experiment, i.e. unitary evolution from steps 1 to 5,
followed by sophisticated
measuring ability in the actions performed by $\overline{W}$ and $W$.
On a single-world interpretation of quantum theory,
this is perfectly acceptable because the interactions carried out in
steps 2 and 5 are not measurements but {\em measurement-like}. They leave
no permanent record, so they must be treated as unitary, following
the Schr\"{o}dinger equation. They contribute to a larger extended event
or history whose probability can be obtained by the Born rule.
This larger extended event or history consists of steps 1--5 and the 
measurements by Wilbur, Wilburbar.

We could also arrange for robot $\overline{F}$ to state her reasoning in the following
way (in the quantum subroutine invoked when she sees tails)
\begin{quote}
(Argument A3): ``I see tails, but I must be careful not to overstate my 
conclusions.
Quantum theory asserts that for a part of the total state
in which my memory, or some other record, unambiguously indicates tails,
the state of the two labs will arrive at $\ket{\psi_5'}$ 
and consequently will lead to an observation of {\em plus} by
Wilbur. However, if this experience of mine is erased or otherwise
modified (as in fact it will be in the protocol I have agreed with
Wilburbar) then other outcomes are possible."
\end{quote}

\subsection{Correct logic in conditions of uncontrolled phases and/or permanent records}

Section \ref{s.Ev} set out how quantum theory and logic
is applied to the protocol described by FR, in which the evolution is unitary
and the phases are under control. We shall now discuss the case
where the experiment is less fully controlled. In this case
whenever a macroscopic system is involved there is liable to occur either
an entanglement with a degree of freedom which is not completely controlled
or contained in the laboratories, or else an interference phase whose value
cannot be controlled or predicted. As Section~\ref{s.project} illustrated,
these two cases lead to the same outcome for the density matrix
(the detailed analysis of such effects is the subject of
decoherence theory). When this is the situation, the total state
of the various systems will develop to $\ket{\tilde{\psi}_5}$, 
not $\ket{\psi_5}$, where we have adopted the convenience of treating
the decoherence by including an uncontrolled phase in the system state.
In this scenario, rational agents will invoke for their reasoning 
whatever follows from the
actual evolution, with its decoherence, not some other evolution.

As explained in Section~\ref{s.project},
the state $\ket{\tilde{\psi}_5}$ is observationally indistinguishable from an
equally weighted probabilistic mixture of $ \ket{\vh\,h,\,\Dn\,\dn}$
and $\ket{\vt\,t,\, \Dn\, \dn}$ and
$\ket{\vt\,t,\, \Up\, \up}$. 
Wilburbar, upon observing {\em minus}, infers that the other lab
is in one of the states $\ket{\vh\,h}$ or $\ket{\vt\,t}$. 
He then infers that Wilbur may observe either {\em plus} or {\em minus}.

\subsection{Other thought-experiments}

Some further thought-experiments have been proposed in recent years,
related in various degrees of directness to the one of FR \cite{Healey2018,Ormrod2022,Brukner2018}. The
common feature is that a process described in words as `measurement' or `observation',
involving an observer who could be a human being, is described mathematically
using a state such as $\ket{{\psi}_5}$ (eqn (\ref{step5})) when it ought to be described using
a state such as $\ket{\tilde{\psi}_5}$ (eqn (\ref{psi5tilde})). That is,
one must keep explicit in the notation the fact that very strong assumptions are
being made about the reproducibility of the phases $\alpha$ and $\overline{\alpha}$
and about the correctness of ignoring information loss (c.f theorem 1 and 
the rest of Section~\ref{s.isolated}). Then, when reasoning about what can be deduced (or asserted without the risk of contradiction) concerning any given scenario, one may correctly present different arguments for the case where the phases are reproducible and the case where they are not. A many-world theory asserts at the outset that all phases are reproducible in principle. A single-world theory asserts that not all phases are reproducible in fact.

In \cite{Kastner2024}, Kastner sets out a refutation of an argument of
Adlam and Rovelli \cite{Adlam2023}. The latter authors intended to show that
Relational Quantum Mechanics (RQM) can consistently account for 
EWF-B thought-experiments; Kastner refutes this, in my view successfully. The
issue here is whether what happens to `inner' observers such as Fiona and 
Fionabar can, on a RQM view, be consistent with what happens to `outer' 
observers such as Wilbur and Wilburbar. Kastner points out that the outer
observer can communicate to the inner observer evidence that they find the latter to be in a superposition.
For example, in our discussion we can append a further two-state system, C to the whole setup, prepared in a state $\ket{0}_{\rm C}$, such that at step 5 
of the protocol with measurement-like interaction the overall state is
\be
\ket{\psi_5} \otimes \ket{0}_{\rm C}.
\ee
An outer observer, whether Wilber, Wilberbar or another, could now measure an 
observable of which $\ket{\psi_5}$ is an eigenstate. This requires them to enact an appropriate interaction involving both laboratories, but this measurement has no effect on the state of the laboratories since the latter are already in an eigenstate of the measured observable. If the outcome of this measurement is consistent with the state $\ket{\psi_5}$ (as we confidently expect it to be) then the system $C$ is made to change its state to $\ket{1}_{\rm C}$. This system can now be sent inside the two laboratories in turn. Further measurement-like interactions between C and some further system in each might
also be performed so as to record this information. When the whole protocol is repeated many times, C is found in state $\ket{1}_{\rm C}$ on every occasion, and this fact is on record inside the laboratories. 

The above kind of communication can serve to make certain claims about the consistency of RQM invalid, argues Kastner. The point for us is to enquire whether this brings anything new to the present argument. It does not.
For, if the interactions are measurement-like then the programming of
robots $F$ and $\overline{F}$ need not change. And if the interactions are measurements then at step 5 we have the state $|\tilde{\psi}_5 \rangle$
(Eqn (\ref{psi5tilde})), so before interactions involving C the overall state is
$
| \tilde{\psi}_5 \rangle \otimes \ket{0}_{\rm C}.
$
Now,
\be
\bra{ \psi_5} \tilde{\psi}_5\rangle = \frac{1}{3}
\left[ e^{i\alpha}   
+ e^{i(\overline{\alpha} + \alpha)}  
+ e^{i\overline{\alpha}} \right].                
\ee
It follows that in this case the probability that a measurement of the pair of laboratories shall yield $\ket{\psi_5}$ varies between 0 and 1, depending on the phases.
Consequently, when the result of such a measurement is communicated to Fiona and Fionabar (via system C), they will not always receive $\ket{1}_C$ and neither
of them will have reason to conclude that their own state was a superposition, or 
otherwise inconsistent with their own observations on a single-world view.

We now turn to a significant no-go theorem arrived at by Bong and co-workers.

Building on work of Brukner, \cite{Brukner2018}, Bong {\em et al.} have obtained a `strong no-go theorem' concerning what can be deduced from correlations among quantities described by them as `measured' or `observed' quantities \cite{Bong2020}. They consider a thought-experiment proposed by Brukner which I shall call EWF-B, to distinguish it from that of FR which is hereafter called EWF-FR.
The no-go theorem is said to be {\em strong} in the sense of ruling out a wide range of physical theories. They state the theorem thus:
\begin{quote}
   ``If a superobserver can perform arbitrary quantum operations on an observer and its environment, then no physical theory can satisfy Local Friendliness.''
\end{quote}
and at the end of their proof, they conclude:
\begin{quote}
   ``In summary, if quantum measurements can be coherently performed
at the level of observers, quantum mechanics predicts the
violation of the LF inequalities in EWFSs. This proves Theorem 1.''
\end{quote}
Here, {\em Local Friendliness} LF is the combination of properties formally defined in their paper and called by them `Absoluteness of Observed Events', 
`No-Superdeterminism' and `Locality'. One would ordinarily expect our experiences to satisfy LF.

The proof offered by Bong {\em et al.} is sound in a mathematical sense, but their terminology, including the statement of the theorem, attaches the term `observer' to a system on which arbitrary quantum operations can be applied. I hold this to be misleading terminology. 
If one characterises as an `observer' a system which can be manipulated to such 
an extent as to achieve quantum erasure of what was said to have been 
`observed', then one adopts misleading terminology. Terminology is agreed
by human convention, but it is best if our conventions are such as to make
scientific language correspond to what is found in the physical behaviour. 
I think that what Bong {\em et al.} show remains valuable, but it should be understood as follows. 

First, we adopt the following:
\begin{quote}
{\bf Definition 1.} {\em the term `observation' shall only be applied to
outcomes which cannot be quantum-erased; the term `observer' shall only
be applied to systems which perform such observations.}
\end{quote} 
Whether the `cannot' mentioned in the definition is because dynamical laws forbid it or because in the event the dynamics make it impossible in practice, we do not need to know. This definition is not one I have seen explicitly stated before, but it is implicit in such works as Peres \cite{Peres1986} and in any exposition of the Copenhagen Interpretation.
Having adopted this approach to terminology, Theorem 1 of Bong {\em et al.} 
has to be stated differently, because the version above now appears to be saying 
``If a self-contradictory physical scenario were to arise, then~\ldots'', which of course does not express their result. A better statement would be
\begin{quote}
If observations satisfy Local Friendliness then if an observer can perform 
arbitrary quantum operations on a given system, the behaviours arising in 
that system do not attain the status of observations, properly so called.
\end{quote}
In the terminology of Definition~1, reversible entanglements in the logical state of a quantum computer do not constitute observation.
The experiences of observers, properly so called by Definition~1, 
can satisfy Local Friendliness without contradicting quantum theory, as far as we know, and the theorem of Bong {\em et al.} does not deny this.

\section{Single-world and many-worlds revisited}  \label{s.singlerev}

In an Everettian treatment of the FR thought-experiment,
Fionabar is unable to make definite predictions based on
any given observation of hers (such as the observation of tails), 
because she has to allow
that in her future all traces of that very observation may be erased and the
universe may proceed as it would if she had observed something else entirely.
Thus a rational agent committed to the relative-state model has to qualify 
their every assertion by the qualification, `as long as future evolution does 
not undo what I have observed, \ldots'. 

In order to be assured that the condition of this qualification is satisfied, 
it is sufficient if a permanent record exists of
the observation in question, or, equivalently, that the process of obtaining
it is `irreversible' in the weak sense of `does not at any future time come
to be reversed' (where the word `reversed' is used in the strong sense of
`leaving no record of any kind'). But a permanent record is the very thing
which makes other ways of interpreting quantum theory legitimate. Broadly 
speaking, single-world quantum theories make the claim that 
the mathematical apparatus of Hilbert space and Schr\"{o}dinger equation does 
not map directly to the physical evolution. Rather, the mathematics captures 
the various influences in such a way as to correctly yield
the probabilities for events and 
histories. In order to support this
claim one requires that there is a way to identify when an event with
probability, as opposed
to unitary evolution with quantum amplitude, is taking place. There need not
be any finite region of spacetime to which one can point and say `here is where the evolution
is not unitary'; rather one takes the view that
in practice there are physical developments which leave permanent records.
Each such
development can be studied by all the usual tools of quantum theory, and
one encounters the type of process treated in decoherence theory. Such
processes lead to a preferred basis. The single-world interpretive 
posture asserts that processes of this kind are what is going on in the
physical universe, and their result is that the universe evolves, such
that the next state of the universe can be modelled mathematically by
one of the states in the preferred basis. 
Quantum theory describes the features which 
lead to the specification of a basis and the Born rule gives
the probabilities for the outcomes.
The physical world arrives at one of those outcomes, not because it follows
Schr\"odinger's equation directly, like a puppet on a string,
but because it evolves according to its own true physical nature, 
and Schr\"odinger's equation captures relevant contributing factors.
Our experience is consistent with this.

The present article is mainly concerned with the distinction between many worlds 
and one world, not with distinguishing among single-world interpretations.
The above is, or can be seen as, a version of the Copenhagen Interpretation (CI) 
or as a modal interpretation or another. Let the reader decide.
It could be seen as a way to describe what it is about the physical system
called `classical apparatus' in CI which underwrites its role in that interpretation. There is also a certain sympathy with the 
Transaction Interpretation, in that I agree the evolution is non-unitary
in fact, even though unitary evolution of the ket plays a large role in the 
analysis; the non-unitarity is asserted by {\em fiat} in the overall statement 
of the structure of the physical theory. 
By making the claim that the physical evolution is non-unitary, we contradict the Everettian 
model, but the two approaches are so constructed that this contradiction cannot 
be empirically demonstrated: one says there are multiple worlds, the other asserts (tautologically and therefore truly) that if another world never,
in the future, has an influence on the one we observe, then that other world is 
empirically irrelevant.

In the taxonomy of interpretations of quantum mechanics proposed by
Cabello,\cite{Cabello2017} the type-II (participatory realist)
interpretations do not propose the projection of a state into a given subspace
(such as an eigenspace of a measured observable) as a physical process that 
starts or stops at any particular time. Rather it is an interpretive
posture concerning what the mathematical statements mean; in particular,
how they furnish information about the developing physical situation.
The notion of `the quantum state of the universe'
is, in such an approach, misleading. Rather we should think of quantum theory as
furnishing correctly the quantum amplitude for any transition from
specified initial
to final conditions, and it is either our physical experience, or our noting
that decoherence and a preferred basis emerge in a calculation, which 
tells us when the modulus-squared of that amplitude can be interpreted
as a probability. On this view, many-world and single-world approaches to quantum theory 
have the same complexity of mathematical apparatus---because it is the
very same apparatus.
The single-world approach has a much simpler ontology.

In applying such a single-world interpretation in practice, one might not know
in some cases whether or not a given development is permanent, and perhaps one
would never know this, just as the Everettian may not. So in either case there
is an implicit qualification attached to any prediction one may make on the
basis of quantum theory.

\subsection{On conscious experience}

The relation between observation and conscious experience is a subtle matter whose thorough treatment is beyond the remit of the current paper. However, for the sake of clarity, I will make a few remarks.

In the case of conscious experience had by humans to date, it is almost certainly  true that all of it is associated with irreversible processes and therefore qualifies as `observation' according to Definition 1. However one naturally wishes to know what might be the situation more generally. 

If a process were to happen so as to bring about a conscious experience, and that process were subsequently reversed, then strictly no physical evidence of that conscious experience would remain, so who is it, at the end, who can be said to have had the experience? Maybe such reversals have happened repeatedly for you and I! If they did then the purported experiences have had no influence whatsoever on who we are now. Therefore, if, in such a case, we point to a body and assert ``the person whose body this is had the experience" then it is debatable whether we speak truly. From this metaphysical puzzle I infer that the position advocated in this paper can reasonably remain agnostic about the relation between consciousness and reversibility. But I think it should also be added that to try to manipulate a conscious being so as to erase their experience is deeply problematic from an ethical point of view.

\section{Conclusion}  \label{s.conclude}

This paper has addressed the EWF-FR thought-experiment and the competing theories of (not just interpretations of) quantum mechanics. 
We also remarked on the
EWF-B thought-experiment of Brukner, with its development by Bong {\em et al.}, and what can be deduced. 

Section \ref{s.FR} pointed out that if one were to program a quantum
computer to go through steps which can be interpreted as reasoning,
then in order to map onto correct reasoning the steps have to be programmed
carefully. In particular, when a quantum computer receives as input a 
physical variable $x$, 
reasoning from one value of $x$ (e.g. the case $x$=`heads')
should not be constructed as if that is the only value that the
quantum computer is handling. An argument such as A1 must not be employed;
the program should be designed to be consistent with an argument such as
A2 or A3.

A single-world quantum mechanics does not deny that something
like a quantum computer could serve the role of $F$ and $\overline{F}$ in
the EWF-FR thought-experiment, and it would be arranged
to calculate consistently, following argument A2 (or A3) not A1. Thus
the contradiction associated with argument A1 is avoided. But
a single-world interpretation may deny that a large system with chaotic
internal dynamics can serve the role of $F$ and $\overline{F}$,
because for such a system it will not be possible for Wilbur and Wilburbar to perform the envisaged measurements.

According to physics
as it stands today, one cannot be certain that some given evolution might not
eventually be reversed and quantum-erased, and this places 
limits on what can be asserted with
confidence in a single-world interpretation. In Section~\ref{s.singlerev} I have 
noted that this fact equally places limits on what can
be asserted with confidence by a rational agent adopting an Everettian
approach. An agent adopting the single-world view is unable to say for certain
`this, and only this, happened; not the superposition'. An agent adopting a 
many-world view is unable to say for certain `my reasoning and its conclusion 
has accounted for all branches of the universe which need to be considered'.
In both cases the possibility that leads to the uncertainty is the possibility
of a future erasure in which different parts of a superposition are made
to interfere. Thus if either agent makes a mistake in their reasoning, 
it is the same mistake that they make. 

Thus, in a relative-state (a.k.a. many-worlds) theory,
an observer cannot know in full what is the current state of some given system, not even one they have just observed, because the observation in any given
branch or world does not reveal what may the case in other branches, and
one may not be able to rule out that the branches will later interfere.
A single-world theory avoids this by associating the term `measurement' with the existence of a permanent record, but now one has the difficulty of not
knowing whether any given decoherence is in fact permanent, which makes
the theory equally hard to apply.

The no-go theorem of Bong {\em et al.} is a theorem concerning systems on which
arbitrary quantum operations can be performed, and therefore it has nothing
to say about other systems. In particular it has nothing to say about
observations which satisfy Definition 1. The quantum foundations literature 
often assumes that to deny that some given quantum operation can be performed 
is to contradict the validity of quantum theory, or to proclaim a restriction on its applicability, but it is not so.
The answer to the question `can quantum mechanics be applied to measuring devices?' is `yes, and the result of such application is that measuring devices undergo irreversible decoherence, and therefore formulations such as Copenhagen or Modal or QBism, among others, can be logically consistent.' 

An interesting development would arise if our most basic theoretical models 
involved irreversibility or indeterminacy in such a way that one could assert,
on the basis of finite evidence, that some 
given record can not be quantum-erased, or that some given interference loop will never close with a well-defined interference phase \cite{Steane2007C}.
There are various hints in contemporary physics and mathematics of such
a development. 
Cosmic expansion leading
to a cosmic event horizon will prevent quantum inference being observable
at the largest scales because the observation requires a loop of wordlines,
but the expansion prevents the loop from closing. 
The theory of Turing machines and algorithms shows that most
sequences of digits are random (in an algorithmic sense)
and most real numbers are non-computable;
in view of this one should not be surprised to find genuine 
randomness in physical behaviour. 
Modern physics also suggests that information 
should itself be seen as a physical quantity, and this may lead one to doubt 
that physical parameters are ever perfectly precise, as Gisin has 
argued \cite{Gisin2017,Gisin2021}. Such imprecision would imply that 
interference phases are not precisely determined. 
Finally, studies of 
quantum gravity and information associated with black holes often suggest information loss and irreversibility,
but the question is not settled.

\backmatter

\bmhead{Acknowledgements}

I acknowledge with thanks the anonymous reviewers whose thorough assessment
and suggestions enabled a significant improvement in clarity and coverage of this work.

\bibliography{./chaosrefs,./philrefs,./myrefs}

\end{document}